\documentclass[cameraready]{Interspeech}

\title{Towards Deep Contextual Reasoning from Broad Descriptions for ASR with Speech-LLM via Metadata-Driven Reasoning Chains}

\author{Jakob}{Poncelet}
\author{Hugo}{Van hamme}
\address{
    KU Leuven \\Department Electrical Engineering ESAT-PSI, Leuven, Belgium
}
\email{}


\usepackage{amsmath,graphicx,hyperref,booktabs,multirow}
\usepackage{subcaption}
\usepackage{enumitem}
\usepackage{pifont}
\usepackage[most]{tcolorbox}
\newtcolorbox{examplebox}{
  colback=black!1,
  colframe=black!25,
  boxrule=0.4pt,
  arc=1.5mm,
  left=1.0mm,right=1.0mm,top=0.7mm,bottom=0.7mm, 
  boxsep=0.6mm,                                   
  before skip=3pt, after skip=3pt,                
}
\newcommand{\exlabel}[1]{\textbf{#1}\hspace{0.5em}}

\begin{document}
\maketitle

\begin{abstract}
Speech recognition often fails on rare, domain-specific terms and context-related named entities. Existing contextualization techniques typically bias decoding with keywords or phrase lists, which does not scale well or exploit deeper knowledge. We propose a training method that teaches a speech-LLM to use broad descriptions (e.g. from videos) as weak semantic priors to perform contextual reasoning grounded in the audio. We build 400 hours of reasoning-augmented speech data by pairing erroneous hypotheses with video metadata and LLM-generated reasoning explanations that justify context-driven corrections. We finetune the speech-LLM to perform chain-of-thought reasoning: generate an initial transcript, then reason over the context, and finally return a corrected transcript. On held-out YouTube-derived test sets, our approach reduces errors, with specific improvements on rare words and named entities, and lays groundwork for deeper contextual reasoning in speech recognition.
\end{abstract}

\section{Introduction} \label{sec:intro}
Current automatic speech recognition (ASR) systems can accurately transcribe in many languages \cite{whisper, owsm, puvvada24_interspeech}, but remain brittle in situations with rare and domain-specific terminology that is acoustically ambiguous. Therefore, research is shifting towards adaptation of pre-trained large language models (LLM) for speech recognition and understanding. In text-based LLMs, leveraging broad context is a core strength: given a topic, their pretrained world knowledge can anticipate plausible entities, disambiguate rare names, and maintain topical consistency. However, modern speech-LLMs and end-to-end ASR systems rarely exploit broad context with the same depth of reasoning routinely observed in text-only LLMs. More specifically, most practical ASR contextualization mechanisms remain quite narrow: they primarily bias the recognizer toward a set of keywords or phrases, rather than enabling the model to reason over a topic description and use that reasoning to guide recognition.

Contextual biasing for end-to-end ASR has advanced through improved training objectives and integration strategies, but the “context” is still typically represented as a bias list (names, terms, hotwords) rather than a broad natural-language description \cite{shakeel24_interspeech,gong25_interspeech}, which is only effective at exact bias word matching. Similarly, speech-LLMs can accept textual prompts and can be pushed toward improved bias-word recognition, yet they can still degrade as the bias list grows and may hallucinate under heavy biasing \cite{gong24b_interspeech}. Parallel lines of work explore injecting richer text into the input \cite{suh24_interspeech} or adding multimedia-derived keywords (e.g., slides) to prompts for LLM-based ASR \cite{yang24f_interspeech}. These results motivate a key question: \textbf{can we adapt a speech-LLM via stepwise chain-of-thought reasoning supervision to more effectively use broad topic descriptions?} 

We propose a method to explicitly teach a speech-LLM to perform deep contextual reasoning from video descriptions that provide weak but informative semantic priors, and to use that reasoning to improve speech recognition while remaining grounded in the acoustics. This grounding is crucial \cite{sun2023contextualbiasing,hu2024listenagain,shivakumar2024speechtext}: in cascaded pipelines that first transcribe speech and then “fix” the transcript with a text-only LLM, the LLM cannot verify whether a proposed correction is acoustically plausible \cite{naderi24_interspeech,zhang25_icassp_rewrites}. By training a speech-LLM that conditions on both the audio and the broad textual context, we aim to obtain topic-consistent corrections that still respect what was actually said.

We construct supervision that links (i) an erroneous transcript, (ii) the available context description, and (iii) a corrected transcript, via an explicit natural-language explanation of why a correction is likely given the context. We then finetune a speech-LLM to produce a stepwise output: first hypothesize a baseline transcript, then reason about the broad context using pretrained world knowledge, and finally apply only those corrections that remain compatible with the acoustic evidence. The use of explicit intermediate reasoning draws inspiration from chain-of-thought prompting and rationale supervision in text models \cite{wei22_neurips,hsieh2023distilling}, but based in a speech-grounded setting.

Our contributions are: (1) a data construction pipeline and open-sourced dataset\footnote{\scriptsize{\url{https://huggingface.co/datasets/kul-speech-lab/contextual-reasoning-speechllm}}} that pairs open-domain speech with real metadata descriptions and produces context-grounded correction rationales, and (2) a training formulation for chain-of-thought ASR that preserves standard ASR behavior while enabling deep contextual corrections from broad descriptions.

\begin{figure*}[t]
    \centering
    \begin{subfigure}[b]{0.428\textwidth}
        \centering
        \includegraphics[height=3.6cm]{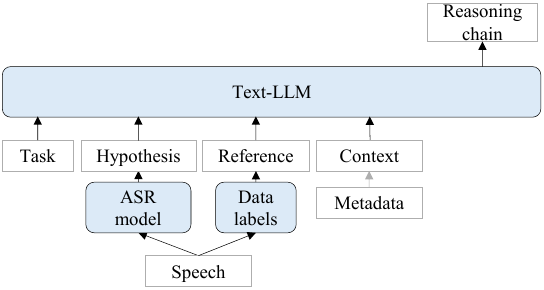}
        \caption{\textbf{Stage 1:} Reasoning chain generation from hypothesis/reference pairs and contextual metadata information.}
        \label{fig:stage1}
    \end{subfigure}
    \hfill
    \rule{0.5pt}{4.5cm}
    \hfill
    \begin{subfigure}[b]{0.53\textwidth}
        \centering
        \includegraphics[height=3.6cm]{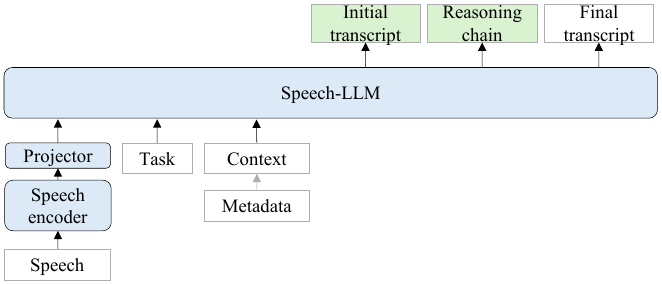}
        \caption{\textbf{Stage 2:} Deep contextual reasoning speech-LLM training/finetuning in chain-of-thought style.}
        \label{fig:stage2}
    \end{subfigure}
    
    \caption{\textbf{Two-stage approach} for speech-LLMs that can leverage broad descriptive information to perform deep contextual reasoning.}
    \label{fig:method}
    \vspace{-10pt}
\end{figure*}

\section{Related Work} \label{sec:related}
\noindent\textbf{Contextualization for ASR.} Contextual ASR typically biases decoding toward a phrase list (entities, catalog items), with improved objectives and scalable integration \cite{shakeel24_interspeech}. In SpeechLLMs, prompt-based bias lists and fusion mechanisms can degrade as lists grow, resulting in hallucinations \cite{gong24b_interspeech}. With retrieval techniques, a small candidate set can be constructed from large vocabularies to alleviate some of these issues \cite{gong25_interspeech}. Finally, richer context injection from metadata or auxiliary media (e.g. slides) helps specialized ASR \cite{suh24_interspeech,yang24f_interspeech} but is often still keyword-centric, instead of implicit deep reasoning from broader descriptions.

\noindent\textbf{LLM post-editing, rescoring, and coupled decoders.} LLM post-editing techniques \cite{zhang25_icassp_rewrites} can yield edits not supported by the acoustics, although techniques like confidence-aware prompting can slightly reduce this risk \cite{naderi24_interspeech}. In contrast, multimodal rescoring and phonetic candidate expansion better preserve audio evidence \cite{shivakumar25_icassp_rescoring,vangysel25_interspeech}. Finally, coupled decoding integrates pretrained LLM decoders during recognition of ASR models \cite{mittal24_interspeech,mittal25_interspeech}. Our work is complementary: rather than focusing on scaling keyword bias lists or purely text-based post-editing, we focus on training a SpeechLLM to transform broad topic descriptions into deep contextual constraints that guide recognition while remaining grounded in the audio.

\noindent\textbf{Reasoning in audio language models.} Recent large audio-language models (LALM) have begun to target explicit reasoning over speech and general sounds, but current evaluations suggest that multi-hop and expert-level reasoning from audio remains difficult \cite{yang25g_interspeech,sakshi2025mmau}.
Several methods therefore introduce reasoning scaffolds, either at inference time via CoT-style prompting \cite{ma2025audiocot}, or through supervised training on structured stepwise reasoning traces \cite{xie2025audioreasoner, kuan25_icassp_stepwise_audio_reasoning}. 
Finally, recent work has shown that textual LLM-generated instruction outputs can be distilled into audio-conditioned models \cite{lu24c_interspeech,Lu2025Developing,lu2025desta25audio,ghosh-etal-2024-gama}.
Rather than reasoning about the audio content, this work uses broad textual metadata to provide topic-level context for ASR. Moreover, these LALM works focus on aspects of speech (such as emotion, speaker, etc) rather than predicting the speech content.

\section{Method} \label{sec:method}
First, we generate reasoning chains from textual data with a text-based LLM, explaining how errors in a transcript can be solved by reasoning about the given context description. Second, we finetune a speech-LLM to perform chain-of-thought error-correcting ASR with these reasoning chains.

\subsection{Reasoning Chain Generation with Video Metadata} \label{sec:dataset}
This section explains the pipeline for the creation of speech reasoning examples and the details of the resulting dataset.
\begin{enumerate}[wide, labelwidth=!, labelindent=0pt]
    \item \textit{\textbf{Data collection:}} We combine transcribed speech data from several open-source datasets with audio derived from YouTube videos: GigaSpeech \cite{chen21o_interspeech} (only the YouTube part of the L set, 975h), SlideSpeech \cite{slidespeech} (L95 set, 473h), SlideAVSR \cite{wang-etal-2024-slideavsr} (train, 29h) and M³AV \cite{chen-etal-2024-m3av} (only used for testing). While Gigaspeech is very broad with diverse topics, SlideSpeech, SlideAVSR and M³AV are more focused on technical and academical presentations with many domain-specific rare words. For segments arising from training splits, we merge consecutive segments such that the average duration is longer ($\sim$7s) than the original segments ($\sim$3s). Merged segments include more contextual information and better match the test sets ($\sim$7s). 

    \item \textit{\textbf{Context generation from metadata:}} For every segment in the dataset, we extract the YouTube URL of the video from which it was derived. Then, we extract video titles, descriptions and tags from the URLs using Google API Client. As these video descriptions are quite noisy with many URLs and hyperlinks, we clean them with an LLM (\textit{Qwen2.5-14B-Instruct-AWQ-4bit}). The prompt specifies instructions to remove all URLs and promotional text, to keep all video content, and to extract entities related to names and topics into a list of tags. The combination of video title, cleaned description, and tags constitutes the context for the speech model.

    \item \textit{\textbf{Pseudolabel/hypothesis generation:}} Our method requires examples of transcripts with mistakes related to the context, such that we can derive reasoning chains for the model to correct them. To get real plausible alternatives as a result of acoustic confusion, we generated pseudolabels with Whisper for all merged segments, using faster-whisper with beam size of 5. More specifically, we use multiple Whisper variants (\textit{large-v3}, \textit{small}, \textit{base}, \textit{tiny}) to have a larger set of more diverse errors (as these models score quite well on the training set). Then, we only retain the segments where the pseudolabels contain an error with respect to the reference transcript that can be attributed to a named entity and/or a rare word. We use \textit{spacy} for named entity recognition from reference transcripts and consider rare words to be not part of the top 5000 most common words in the whole dataset. Next, we throw out all ``audio-only" errors, such as trivial errors to function words, derivations, inflections and pronouns, that are not related to context but only to the audio.

    Finally, we also create an additional set of LLM-generated artificial errors by using a prompt that carefully instructs to insert acoustically plausible errors into the transcript for named entities, keywords or rare words, only if those errors can be resolved via the context (e.g. explicit mention of a name in the description or tags, or a word related to the topic in the description). For this step, we use \textit{Qwen2.5-32B-Instruct-AWQ-4bit}.

    \item \textit{\textbf{Reasoning chain generation:}} We prompted a text-based LLM with the reference transcript, an erroneous hypothesis, a pre-made list of corrections (i.e. WER alignment), the cleaned video context (title, description and tags), and for every correction whether the target word is explicitly in the context. The LLM is instructed to generate a reasoning chain or explanation for each correction from hypothesis to reference based on the context. The concept is depicted in Figure~\ref{fig:stage1}.
    
    For target words that are explicitly mentioned in the context, the reasoning chain must mention so. For other errors, the LLM should attempt to perform deep reasoning. It should cite specific cues from the context and explain how they make the specific correction probable (without mentioning the reference), or alternatively, decide that the correction is not derivable from contextual information. For example, if the description is about sailing in Sweden, then Swedish town names are more likely to appear as well as nautical terms like 'buoy'. If the video is about rock bands, then the specific name of a band manager is more likely. Finally, the LLM should output a JSON format with explanations for every correction.
    We use \textit{Qwen2.5-32B-Instruct-AWQ-4bit} to generate the structured reasoning chains. Figure~\ref{fig:example} shows an example of a hypothesis and reference and a generated reasoning chain leveraging the video description.
\end{enumerate}

\begin{figure}[t]
    \centering
    \begin{examplebox}
    \footnotesize
    \setlength{\parskip}{0pt}
    \setlength{\parindent}{0pt}
    \exlabel{Reference:} \textit{[...] prevented him from integrating into the \textbf{jinling}} \\
    \exlabel{Hypothesis:} \textit{[...] prevented him from integrating into the \textbf{jingling}} \\
    \exlabel{Reasoning:} \textit{In the hypothesis, 'jingling' is a phonetic error for 'jinling', which is a plausible correction given the context of the presentation on Chan Temple and the historical period discussed. The term 'Jinling' refers to the region around Nanjing, which is relevant to the historical and cultural context of the discussion.}
    \end{examplebox}
    \caption{\textbf{Example} of a generated reasoning chain from a reference transcript, an ASR hypothesis and context (omitted).}
    \label{fig:example}
    \vspace{-10pt}
\end{figure} 

We end up with a total of 400 hours of reasoning-augmented speech examples. For the resulting hypotheses, 75\% of the errors are ASR-based, and 25\% are LLM-generated. The data distributions are detailed in Table~\ref{tab:data1}. 
Finally, we also derive 2 subsets. The first subset is filtered by an LLM (Qwen2.5-32B) to discard all samples where the reasoning is too far-fetched or incorrect. The second subset is the smallest and only contains segments with ASR-based named entity errors from GigaSpeech. Table~\ref{tab:data2} depicts the details of these sets (L, M and S) and the remaining number of reasoning examples.

For evaluation, we use the (held-out) M³AV data set. As this set is quite large (16k examples), we select all examples containing at least one named entity, leading to 3.9k examples or 9 hours of speech. We extract named entities from the uncased reference transcript using a finetuned BERT \cite{bert} base model.

\begin{table}[t]
    \centering
    \footnotesize
    \caption{\textbf{Reasoning-augmented dataset:} Details of the resulting dataset and splits with speech/context reasoning examples.}
    \label{tab:data}
    \vspace{-5pt}
    \begin{subtable}[t]{\columnwidth}
        \resizebox{\columnwidth}{!}{%
        \begin{tabular}{c|cc c c c}
            \toprule
            \textbf{Dataset} & \multicolumn{2}{c}{\textbf{\# Reasoning examples}} & \textbf{\# Audio} & \textbf{\# Audio} & \textbf{\# Video} \\
             \textbf{source} & \textbf{ASR-error} & \textbf{LLM-error} & \textbf{segments} & \textbf{hours} & \textbf{files} \\
            \midrule
            \textbf{\textit{GigaSpeech}} & 178k & 46k & 108k & 285h & 5.7k \\
            \textbf{\textit{SlideSpeech}} & 46k & 19k & 39k & 107h & 1.3k \\
            \textbf{\textit{SlideAVSR}} & 8k & 3k & 5k & 11h & 0.2k \\
            \bottomrule
        \end{tabular}
        }
        \caption{Training data distribution}
        \label{tab:data1}
    \end{subtable}

    \begin{subtable}[t]{\columnwidth}
        \resizebox{\columnwidth}{!}{%
        \begin{tabular}{c|c|c c c c}
            \toprule
            \textbf{Set} & \textbf{Hypothesis} & \textbf{\# Reasoning} & \textbf{\# Audio} & \textbf{\# Audio} & \textbf{\# Video} \\
            \textbf{id} & \textbf{error source} & \textbf{examples} & \textbf{segments} & \textbf{hours} & \textbf{files} \\
            \midrule
            \textbf{\textit{L}} & ASR+LLM & 300k & 152k & 403h & 7.2k\\
            \textbf{\textit{M}} & ASR+LLM (filt.) & 130k & 86k & 232h & 6.8k \\
            \textbf{\textit{S}} & ASR (NE only) & 27k & 12k & 41h & 2.8k \\
            \bottomrule
        \end{tabular}
        }
        \caption{Final training data splits}
        \label{tab:data2}    
    \end{subtable}
    \vspace{-20pt}
\end{table}

\subsection{Deep Contextual Reasoning with Speech-LLM}
Generally, speech-LLMs are (pre-)trained on ASR data to output a transcript for a speech input. To maintain consistency with pre-training behaviour and allow joint finetuning with ASR data, we teach the LLM to perform contextual reasoning in a chain-of-thought manner \cite{wei22_neurips}. The speech-LLM first has to return an initial transcript of the speech, and then use the context for reasoning and error correction. So, the speech-LLM is finetuned to output the following format, as depicted in Figure~\ref{fig:stage2}:

\noindent \verb|<initial-text>| - \verb|<reasoning>| - \verb|<final-text>|

The initial transcripts are derived from the pseudolabels, the reference and the reasoning chains in a reverse manner: for every correction from pseudolabel to reference that is justified by context according to the LLM-generated reasoning chain, apply the reverse correction to the reference. So conceptually, only those errors that can be corrected based on context are present in the pseudolabel. If the pseudolabel itself is used as initial text, there are often many audio-only errors. This results in teaching the model to make many corrections to its initial transcript, and additionally, can lead to a train-test mismatch when the speech-LLM becomes better than the pseudolabeling model. As the scope of this paper is contextual biasing, we do not focus on these audio-only or grammatical spelling errors here.

Finally, during training, we combine pure ASR data with context-reasoning data (50/50 division per batch). For the ASR data, the reasoning chain collapses to "No contextual errors" and the reference transcript is used as initial transcript. Hence, the model learns that it doesn't always have to make corrections to the initial transcript. Note that in all cases, the prediction loss on the initial transcript is masked, and only the loss on the reasoning chain and final transcript are incurred, such that the model does not learn to mimic the pseudolabels.

We use speech-LLMs with 4-bit LLM quantization and finetune QLoRA blocks \cite{qlora} on all linear layers in the LLM except the prediction head, with rank 32 (or 16 when training on the S set), alpha 64 and dropout 0.05. The speech encoder and projector are not adapted. This gives 40M (vs 20M) trainable parameters for QLoRA rank 32 (vs 16).
The models are finetuned for 1-5 epochs (depending on set size) until convergence with an effective batch size of 128, using the 8bit Adam optimizer with a linear decaying learning rate and peak value of 1e-4, 100 warmup steps, and 0.1 weight decay.
With a context window of at least 8192 tokens in the speech-LLMs, we never exceed the limit, as the longest description is only 3,000 tokens.

\section{Experimental Results} 

\subsection{Reasoning with Qwen2-Audio}
We finetune the Qwen2-Audio-7B-Instruct speech-LLM on the three dataset splits and evaluate on the diverse and challenging M³AV test set. We report several baselines: 1) the non-finetuned model, 2) finetuning with plain ASR/transcribe task without context, 3) finetuning with contextualized transcribe task (i.e. ``use the context"), and 4) finetuning with a two-stage prediction pipeline using context, similar to the proposed method but without the explicit reasoning chain. Results are in Table~\ref{tab:exp1}, reporting global, rare, and named-entity based WER.

\begin{table}[t]
    \footnotesize
    \centering
    \caption{\textbf{Qwen2-Audio:} WER (\%) on M³AV test set when finetuning Qwen2-Audio-7B on different splits. We report global WER (all), rare word WER (rare), and named entity WER (NE).}
    \label{tab:exp1}
    \vspace{-7pt}
    \resizebox{\columnwidth}{!}{%
    \begin{tabular}{c | c | c c c}
        \toprule
        \textbf{Train} & \textbf{Prompting} & \multicolumn{3}{c}{\textbf{WER (\%)}} \\
        \textbf{set} & \textbf{method} & \textit{All [83k]} & \textit{Rare [13k]} & \textit{NE [7k]} \\
         \midrule
         \multirow{2}{*}{-} & Transcribe & 13.1 & 30.0 & 28.9 \\
         & Transcribe w/ Context & 257.6 & 69.2 & 68.3 \\
         \hline
         \multirow{4}{*}{\textbf{\textit{S}}} & Transcribe & 11.9 & 31.2 & 29.6 \\
         & Transcribe w/ Context & \textbf{11.0} & 27.9 & 26.9 \\
         & 2-stage Transcribe & 11.5 & 28.2 & 27.4 \\
         & 2-stage Reason (ours) & \textbf{11.0} & \textbf{26.3} & \textbf{26.1} \\
         \hline
         \multirow{4}{*}{\textbf{\textit{M}}} & Transcribe & 10.2 & 27.2 & 26.4 \\
         & Transcribe w/ Context & 9.8 & 24.2 & 23.8 \\
         & 2-stage Transcribe & 9.4 & 24.0 & 23.9 \\
         & 2-stage Reason (ours) & \textbf{9.3} & \textbf{23.1} & \textbf{23.3} \\
         \hline
         \multirow{4}{*}{\textbf{\textit{L}}} & Transcribe & 11.1 & 28.3 & 27.2 \\
         & Transcribe w/ Context & 10.4 & 24.8 & 24.7 \\
         & 2-stage Transcribe & 9.9 & 25.0 & 24.5 \\
         & 2-stage Reason (ours) & \textbf{9.5} & \textbf{23.4} & \textbf{23.6} \\
         \bottomrule
    \end{tabular}
    }
    \vspace{-10pt}
\end{table}

The base model is not able to use large text contexts and dramatically hallucinates. On the other hand, when finetuned with context and explicitly taught to use that context, the recognition of standard and specialized words improves (this is similar to standard keyword biasing methods). Our proposed two-stage reasoning approach further improves recognition, especially for rare words and named entities which it can derive from context. Note that reasoning is a much more difficult task to learn with the same parameter count. The filtered M set seems to instill better reasoning chains, although the L set would probably benefit from further tuning of training hyperparameters.

\subsection{Contextual Reasoning Analysis}
First, we evaluate the difference in WER for named entities and rare words between the initial hypothesis transcript and the final transcript in our speech-LLM reasoning method, using the reasoning models from Table~\ref{tab:exp1}. We also correct the same transcripts with a text-based LLM, with or without context. We use \textit{Qwen2-7B-Instruct} as it is the same backbone LLM used in \textit{Qwen2-Audio}. We also report the percentage of reasoning chains that positively/negatively impacted WER, in Table~\ref{tab:exp2}.

\begin{table}[t]
    \footnotesize
    \centering
    \caption{\textbf{Error correction analysis:} Comparing initial to final predictions of reasoning models from Table~\ref{tab:exp1} (M³AV test set). Text-only LLM corrections with/without context (Qwen2-7B) are reported too. We report global, rare, and named entity (NE) WER. The last four columns report the percentage of examples for which a reasoning correction was made (\%), and the percentage of those corrections that had a positive (POS), neutral (EQ) or negative (NEG) impact on WER (positive=reduction).}
    \label{tab:exp2}
    \vspace{-7pt}
    \resizebox{\columnwidth}{!}{%
    \begin{tabular}{c | c | c c c | c c c c}
        \toprule
        \multirow{2}{*}{\textbf{Set}} & \multirow{2}{*}{\textbf{Predicted transcript}} & \multicolumn{3}{c|}{\textbf{WER (\%)}} & \multicolumn{4}{c}{\textbf{Corrections}} \\
         & & \textit{All} & \textit{Rare} & \textit{NE} & \textit{Corr. \%} & \textit{POS} & \textit{EQ} & \textit{NEG} \\
         \midrule
         \multirow{4}{*}{\textbf{\textit{S}}} & Speech-LLM initial transcript & 11.4 & 28.4 & 27.3 & -- & -- & -- & -- \\
         & Speech-LLM correction (reason) & \textbf{11.0} & \textbf{26.3} & \textbf{26.1} & 43.6 & 35.1 & 42.2 & 22.6 \\
         & Text-LLM correction (no context) & 15.4 & 27.3 & 26.5 & -- & -- & -- & -- \\
         & Text-LLM correction (with context) & 20.1 & 28.3 & 27.4 & -- & -- & -- & -- \\
         \hline
         \multirow{4}{*}{\textbf{\textit{M}}} & Speech-LLM initial transcript & 9.6 & 24.6 & 24.2 & -- & -- & -- & -- \\
         & Speech-LLM correction (reason) & \textbf{9.3} & \textbf{23.1} & \textbf{23.3} & 29.4 & 31.5 & 51.9 & 16.5 \\
         & Text-LLM correction (no context) & 14.3 & 25.2 & 26.5 & -- & -- & -- & -- \\
         & Text-LLM correction (with context) & 19.8 & 26.7 & 25.8 & -- & -- & -- & -- \\
         \bottomrule
    \end{tabular}
    }
    \vspace{-10pt}
\end{table}

Text-LLMs have a lot of difficulty sticking to the original transcript, even when explicitly instructed, and report higher WERs than the original transcript. This could be improved by feeding N-best lists, which was not implemented in Qwen2-Audio. Moreover, the results show that reasoning effectively leads to grounded transcript corrections with significant drops in WER. While the reasoning chains are not perfect yet, the trend is positive and might further benefit from larger models.

\subsection{Impact of Pre-trained Model}
We evaluate the proposed method on alternative speech-LLMs and assess the impact, since \textit{Qwen2-Audio} might have lost some of its textual knowledge and reasoning capabilities as a result of large-scale training on audio data. We compare: 1) \textit{Qwen2.5-Omni-7B}, a multi-modal model trained for text/image/speech understanding and text/speech generation, 2) \textit{Audio-Flamingo-3} (7B), an audio reasoning model, 3) \textit{Ultravox-v0.5-8B}, a speech-LLM built around a completely frozen \textit{Llama-3.1-8B} backbone. Results are in Table~\ref{tab:exp3}. 
We do not find major differences between  model types, but we still observe improvements through our reasoning method across all models and training splits.

\begin{table}[t]
    \footnotesize
    \centering
    \caption{\textbf{Model comparison:} WER (\%) on M³AV test set. The models are finetuned to transcribe (ASR), transcribe with context (C-ASR), or to do two-stage reasoning with context (Reason). We report global, rare and named entity (NE) WER.}
    \label{tab:exp3}
    \vspace{-7pt}
    \resizebox{\columnwidth}{!}{%
    \begin{tabular}{c | c | c c c | c c c | c c c}
        \toprule
        \multirow{2}{*}{\shortstack{\textbf{Train}\\\textbf{set}}}& \multirow{2}{*}{\shortstack{\textbf{Prompt}\\\textbf{method}}} & \multicolumn{3}{c}{\underline{\textit{\textbf{Qwen2.5-Omni-7B}}}} & \multicolumn{3}{c}{\underline{\textit{\textbf{Audio-Flamingo-3}}}} & \multicolumn{3}{c}{\underline{\textit{\textbf{Ultravox-V0.5-8B}}}} \\
         & & \textit{All} & \textit{Rare} & \textit{NE} & \textit{All} & \textit{Rare} & \textit{NE} & \textit{All} & \textit{Rare} & \textit{NE} \\
         \midrule
         \multirow{2}{*}{-} & ASR & 12.8 & 20.8 & 22.3 & 8.6 & 20.4 & 21.2 & 27.9 & 25.3 & 26.1 \\
         & C-ASR & 28.8 & 23.7 & 24.9 & 12.9 & 22.2 & 23.0 & 98.4 & 25.5 & 27.1 \\
         \hline
         \multirow{3}{*}{\textbf{\textit{S}}} & ASR & 8.5 & 21.0 & 21.9 & 109.4 & 85.6 & 84.5 & 9.6 & 24.4 & 25.6 \\
         & C-ASR & 8.3 & 19.9 & 20.7 & 101.3 & 87.5 & 88.8 & \textbf{9.2} & 22.4 & 23.3 \\
         & Reason & \textbf{8.2} & \textbf{18.7} & \textbf{20.0} & \textbf{37.1} & \textbf{40.4} & \textbf{48.1} & \textbf{9.2} & \textbf{21.1} & \textbf{22.4} \\
         \hline
         \multirow{3}{*}{\textbf{\textit{M}}} & ASR & 8.0 & 20.1 & 20.9 & 9.7 & 23.0 & 23.6 & 8.4 & 21.1 & 23.1 \\
         & C-ASR & 7.5 & 17.7 & 19.2 & 8.5 & 20.1 & 21.1 & \textbf{7.6} & \textbf{18.4} & 20.4 \\
         & Reason & \textbf{7.3} & \textbf{17.4} & \textbf{18.3} & \textbf{7.9} & \textbf{18.3} & \textbf{19.2} & \textbf{7.6} & \textbf{18.4} & \textbf{20.2} \\
         \hline
         \multirow{3}{*}{\textbf{\textit{L}}} & ASR & 8.9 & 22.1 & 22.6 & 8.9 & 22.7 & 24.1 & 9.1 & 21.3 & 22.9 \\
         & C-ASR & 8.4 & 19.9 & 19.9 & 8.5 & 20.5 & 22.2 & 8.1 & 18.8 & 20.0 \\
         & Reason & \textbf{7.9} & \textbf{19.0} & \textbf{19.8} & \textbf{7.9} & \textbf{18.7} & \textbf{19.7} & \textbf{7.7} & \textbf{18.4} & \textbf{19.9} \\
         \bottomrule
    \end{tabular}
    }
\end{table}

\subsection{Other Evaluation Sets}
We also evaluate on the official test sets from SlideSpeech (8 hours, 3.2k examples) and SlideAVSR (4 hours, 2.1k examples). Results for all models (trained on M split) are in Table~\ref{tab:exp4}.
While there are less context-related rare keywords in these test sets, we still observe  improvements for most models through reasoning, effectively leveraging the video descriptions to bias and improve the prediction.

\begin{table}[t]
    \footnotesize
    \centering
    \caption{\textbf{Alternative test sets:} WER (\%) on SlideSpeech and SlideAVSR test sets with models from Table~\ref{tab:exp3} (M split). The models were finetuned to transcribe (ASR), transcribe with context (C-ASR), or to do two-stage reasoning with context (Reason). We also add non-finetuned results (--). We report global, rare and named entity (NE) WER.}
    \label{tab:exp4}
    \vspace{-7pt}
    \resizebox{\columnwidth}{!}{%
    \begin{tabular}{c | c | c c c | c c c}
        \toprule
        \multirow{3}{*}{\textbf{Model}} & \multirow{3}{*}{\shortstack{\textbf{Prompt}\\\textbf{train}\\\textbf{method}}} & \multicolumn{3}{c}{\underline{\textit{\textbf{SlideSpeech}}}} & \multicolumn{3}{c}{\underline{\textit{\textbf{SlideAVSR}}}} \\
         & & \textit{All} & \textit{Rare} & \textit{NE} & \textit{All} & \textit{Rare} & \textit{NE} \\
         & & \textit{[83k]} & \textit{[6k]} &  \textit{[1.7k]}& \textit{[38k]} &  \textit{[5.9k]} &  \textit{[0.7k]} \\
         \midrule
         \multirow{4}{*}{\shortstack{\textit{Qwen2-}\\\textit{Audio-7B}}} & -- & 10.5 & 17.1 & 30.2 & 16.8 & 29.4 & 53.7 \\
         & ASR & 7.8 & 14.7 & 27.5 & 13.5 & 24.7 & 46.2 \\
         & C-ASR & 8.0 & 13.0 & 25.1 & 12.6 & 21.1 & 37.9 \\
         & Reason & \textbf{7.7} & \textbf{12.6} & \textbf{24.6} & \textbf{12.5} & \textbf{20.6} & \textbf{37.4} \\
         \hline
         \multirow{4}{*}{\shortstack{\textit{Qwen2.5-}\\\textit{Omni-7B}}} & -- & 14.6 & 14.5 & 28.8 & 21.5 & 25.3 & 43.9 \\
         & ASR & 7.4 & 12.4 & 25.4 & 10.0 & 17.9 & 35.5 \\
         & C-ASR & 7.3 & 11.3 & 23.0 & \textbf{9.5} & \textbf{16.0} & 30.8 \\
         & Reason & \textbf{7.2} & \textbf{11.2} & \textbf{22.5} & 9.7 & 16.2 & \textbf{29.9} \\
         \hline
         \multirow{4}{*}{\shortstack{\textit{Audio-}\\\textit{Flamingo-}\\\textit{3}}} & -- & 8.2 & 12.0 & 24.3 & 11.4 & 22.4 & 43.3 \\
         & ASR & 8.5 & 13.1 & 26.6 & 12.8 & 22.1 & 43.6 \\
         & C-ASR & 8.0 & \textbf{11.8} & 24.0 & 12.0 & 19.5 & 37.0 \\
         & Reason & \textbf{7.7} & \textbf{11.8} & \textbf{23.0} & \textbf{11.2} & \textbf{18.5} & \textbf{34.9} \\
         \hline
         \multirow{4}{*}{\shortstack{\textit{Ultravox-}\\\textit{v0.5-8B}}} & -- & 29.2 & 16.6 & 30.2 & 36.5 & 27.8 & 49.1 \\
         & ASR & 8.1 & 13.1 & 26.6 & 10.9 & 20.2 & 41.0 \\
         & C-ASR & 7.7 & 11.8 & 23.6 & \textbf{10.4} & 17.6 & 34.7 \\
         & Reason & \textbf{7.3} & \textbf{11.4} & \textbf{23.0} & 10.6 & \textbf{17.1} & \textbf{32.4} \\
         \bottomrule
    \end{tabular}
    }
    \vspace{-10pt}
\end{table}

\section{Discussion}
The context descriptions used in this work are rather broad compared to time-aligned keywords or slides. Hence, we should not expect massive WER improvements, especially when segments are rather short. Still, we show that chain-of-thought reasoning is possible (by only adapting the LLM) and that reasoning-guided training from such priors is beneficial for speech recognition with topic-related descriptions, which might translate to other domains such as social media video captions, and inspire further work on reasoning speech-LLMs.
There might be also an interesting synergy with our method when adapting even larger thinking or reasoning-based LLMs to speech, yet we were constrained by resources to evaluate this beyond 8B models.

\section{Conclusion}
We proposed a chain-of-thought ASR finetuning method that teaches a speech-LLM to use broad video descriptions for deep contextualized reasoning while staying grounded in the audio. We also introduced a pipeline and dataset that pair metadata with contextual transcript errors and correction rationales. Experiments show improved recognition overall, with the clearest gains on rare words and named entities.

\newpage
\section{Acknowledgments}
\ifcameraready
This research was supported by the Research Foundation Flanders (FWO) under grant S004923N of the SBO programme and by the Flemish Government under the "Flanders AI Research Program". Part of the resources and services used in this work were provided by the VSC (Flemish Supercomputer Center), funded by the Research Foundation Flanders (FWO) and the Flemish Government.
\else
Anonymous.
\fi

\section{Use of Generative AI Disclosure}
The authors used Generative AI exclusively for text formatting, editing, and polishing to improve the clarity of this manuscript. No part of the manuscript's content or ideas was produced by generative AI. All authors take full responsibility and accountability for the original work and content of this paper.

\bibliographystyle{IEEEtran}
\bibliography{refs}

\end{document}